\documentclass[useAMS,usenatbib]{mn2e}
\usepackage[latin1]{inputenc}
\usepackage[english]{babel}
\usepackage{latexsym}
\usepackage{graphicx}

\title[Low redshift constraints on scalar-tensor theories]{Low redshift constraints on scalar-tensor theories}
\author[Stéphane Fay]{Stéphane Fay\footnote{steph.fay@gmail.com}\\
Palais de la Découverte\\
Astronomy Department\\
Avenue Franklin Roosevelt\\
75008 Paris\\
France}
\begin{document}
\maketitle
\begin{abstract}
Scalar-tensor theories are constrained with lunar laser ranging and supernovae data at low redshift. This allows to find some constraints on the scalar field independently on the form of its potential once the gravitation function is specified. We apply these results to some well known scalar-tensor theories showing that they agreed with the $\Lambda$CDM model at $1\sigma$.
\end{abstract}
%------------------------------------------------------------------------------------------------------------------------------------------------------------------------------------------%
\section{Introduction} \label{s0}
Scalar-tensor theories\citep{BraDic61} are among the theories able to explain the Universe late time accelerated expansion\citep{Rie98}\citep{Per99}. They generalise the $\Lambda$CDM or quintessence models by assuming a varying gravitation function $G$. Consequently they can be difficult to constrain because one has to interpret observations, in particular cosmological ones, by considering the variation of $G$. Among the various experiments\citep{Chi11} attempting to constraint $G$, lunar laser ranging \citep{Wil04} gives some stringent limits on the variation of $dG/dt G^{-1}$ today. Another constraint comes from supernovae data whose magnitude-redshift relation have to take into account the variation of $G$ on the supernovae mechanism\citep{RiaUza02}.\\
In this paper we use lunar laser ranging and supernovae data at low redshift ($z<0.1$) to constrain scalar-tensor theories. We show that these two kinds of observations are complementary. Low redshift data allow to consider power series of the field equations near $z=0$. This leads to constrain observationally some of the terms of these series without specifying the form of the scalar field potential $U$. It is then also possible to derive some constraints on the free parameters of a scalar-tensor theory once $U$ is given. In particular, we will choose to consider the theories defined by $G^{-1}=\phi$ with $U=\Lambda$, $1/2m^2\phi^2$ and $me^{\Lambda\phi}$, $\Lambda$ and $m$ being the free constant parameters.\\
The plan of the paper is as follows. In the first section, we write the scalar-tensor theory field equations in two ways. The first one depends on $G$ and its derivatives and allows to use the lunar laser ranging. The second one consists in rewriting the field equations as General Relativity with a perfect fluid modeling the scalar field and defined by a density $\rho$, a pressure $p$ and an equation of state $w$. It is more appropriate to use the supernovae data. Developing some functions (scalar field, Hubble function, etc) as power series of the redshift near $z=0$, we express some of their constant coefficients with observational quantities. In the second section, we use lunar laser ranging and supernovae data to constrain these coefficients. In a third section, these results are discussed and applied to some classes of scalar-tensor theories.
%------------------------------------------------------------------------------------------------------------------------------------------------------------------------------------------%
\section{Field equations} \label{s1}
The scalar-tensor theory action writes
$$
S=\int\left[G^{-1}R-\frac{\omega}{\phi}\phi_{,\mu}\phi^{,\mu}-U+ 16\pi L_m\right]
$$
$\phi$ is the scalar field, $G(\phi)$ the gravitation function, $\omega(\phi)$ the Brans-Dicke function and $U(\phi)$ the scalar field potential. $L_m$ is the matter Lagrangian. We will choose units such that $G(t_0)=G_0=1$ with $t_0$ the proper time today. Varying the action with respect to the metric function $g_{\mu\nu}$, we get the field equations for a flat Universe:
\begin{eqnarray*}
R_{\mu\nu}-\frac{1}{2}g_{\mu\nu}R&=&G(\frac{\omega}{\phi}\phi_{,\mu}\phi_{,\nu}-\frac{\omega}{2\phi}\phi_{,\lambda}^{,\lambda}g_{\mu\nu}+G^{-1}_{,\mu;\nu}-\\
&&g_{\mu\nu}\Box (G^{-1})-\frac{1}{2}Ug_{\mu\nu}+8\pi T_{\mu\nu})\\
\end{eqnarray*}
We want to rewrite them as General Relativity with a perfect fluid\citep{Cap06}. For this, we transform the above equations as
\begin{eqnarray*}
R_{\mu\nu}-\frac{1}{2}g_{\mu\nu}R&=&-(G^{-1}-1)(R_{\mu\nu}-\frac{1}{2}g_{\mu\nu}R)+\frac{\omega}{\phi}\phi_{,\mu}\phi_{,\nu}-\\
&&\frac{\omega}{2\phi}\phi_{,\lambda}\phi^{,\lambda}g_{\mu\nu}+G^{-1}_{,\mu;\nu}-g_{\mu\nu}\Box (G^{-1})-\\
&&\frac{1}{2}Ug_{\mu\nu}+8\pi T_{\mu\nu}\\
\end{eqnarray*}
and we define the impulsion-energy tensor of the perfect fluid as
\begin{eqnarray*}
T_{\mu\nu(\phi)}&=&-(G^{-1}-1)(R_{\mu\nu}-\frac{1}{2}g_{\mu\nu}R)+\frac{\omega}{\phi}\phi_{,\mu}\phi_{,\nu}-\\
&&\frac{\omega}{2\phi}\phi_{,\lambda}\phi^{,\lambda}g_{\mu\nu}+G^{-1}_{,\mu;\nu}-g_{\mu\nu}\Box (G^{-1})-\frac{1}{2}Ug_{\mu\nu}\\
\end{eqnarray*}
We checked that $T_{\mu\nu(\phi)}$ is conserved by calculating $T^{\mu\nu}_{(\phi);\mu}$. We then recover the Klein-Gordon equation that is usually obtained by varying the action with respect to $\phi$. We define the density $\rho$ and pressure $p$ associated to $T_{\mu\nu(\phi)}$ as
\begin{equation}\label{rhoG}
\rho=T_{00(\phi)}=-3(G^{-1}-1)H^2+\frac{1}{2}\frac{\omega}{\phi}\dot\phi^2-3H(G^{-1})^.+\frac{1}{2}U
\end{equation}
\begin{eqnarray*}
p&=&T_{ii(\phi)}/a^2=(G^{-1}-1)(2\dot H+3H^2)+\frac{1}{2}\frac{\omega}{\phi}\dot\phi^2+\\
&&(G^{-1})^{..}+2H(G^{-1})^.-\frac{1}{2}U\\
\end{eqnarray*}
where a dot means a derivative with respect to the proper time $t$. Hence, the equation of state $w=p/\rho$ is
$$
w+1=\frac{-H(G^{-1})^.+2(G^{-1}-1)\dot H+(G^{-1})^{..}+\frac{\omega}{\phi}\dot\phi^2}{-3(G^{-1}-1)H^2+\frac{1}{2}\frac{\omega}{\phi}\dot\phi^2-3H(G^{-1})^.+\frac{1}{2}U}
$$
This last expression agrees with the conservation of $T_{\mu\nu(\phi)}$ since it checks $\dot\rho=-3H\rho(1+w)$.\\\\
In the rest of the paper, we redefine the scalar field without loss of generality such as the kinetic term $\frac{\omega}{\phi}\phi_{,\mu}\phi^{,\mu}$ is cast into $\frac{1}{2}\phi_{,\mu}\phi^{,\mu}$. We choose to study the class of theories defined by $G^{-1}=\phi$, leaving $U$ unspecified.\\\\
Around $z=0$, we write the density $\rho$ and the equation of state $w$ as some power series $\rho=\rho_0+\rho_1 z+O(2)$ and $w=w_0+w_1 z+O(2)$. Then, introducing these series in the energy conservation equation for $\rho$, we get for the zeroth order term
$$
\rho_1=3(1+w_0)\rho_0
$$
It is possible to get equations for any higher order terms but they introduce some constant coefficients $\rho_n$ and $w_n$ that we cannot constrain observationally. So we do not consider them here. Using this expression for $\rho_1$ in the constraint equation
$$
H^2=H_0^2\left[\Omega_{m0}(1+z)^3+(1-\Omega_{m0})\frac{\rho}{\rho_0}\right]
$$
with $1-\Omega_{m0}=\Omega_{0}=\frac{\rho_0}{H_0^2}$, and writing the Hubble function as $H=H_0+H_1 z+O(2)$, we get for the first order term (linear in $z$, the zeroth order term giving the usual constraint $\Omega_{m0}+\Omega_{0}=1$)
\begin{equation}\label{H1}
H_1=-\frac{1}{2}H_0\left[-3+3w_0(\Omega_{m0}-1)\right]
\end{equation}
Developing $\phi$ as $\phi=\phi_0+\phi_1 z+\phi^2 z^2+O(3)$, it comes $\phi_0=1$ since we choose $G_0=1$. Moreover, from the definition (\ref{rhoG}) for the density $\rho$, we get for $\phi_1$ and $\phi_2$ considering zeroth and first order terms
\begin{equation}\label{phi1}
\phi_1=-6\mp\frac{\sqrt{Q}}{H_0}
\end{equation}
\begin{eqnarray}
\phi_2&=&\frac{216 H_0^4\pm\sqrt{Q} \left[ 36 H_0^3+H_1 (2 \rho_0-U(1))\right]}{H_0^2Q}+\nonumber\\
&&\frac{H_0 \left(2 \rho_0-\rho_1-U(1)-3 U_\phi(1)\right)}{H_0^2Q}+\label{phi2}\\
&&\frac{6 H_0^2 \left(4 \rho_0-2 U(1)-3 U_\phi(1)\right)-(2 \rho_0-U(1)) U_\phi(1)}{H_0^2Q}\nonumber
\end{eqnarray}
with $U(1)=U(\phi=1)$, the present value of the potential, $U_\phi(1)=\frac{dU}{d\phi}(\phi=1)$ and $Q=36 H_0^2+4 \rho_0-2U(1)$. In the next section, we constrain $\phi_1$ and $w_0$ thanks to lunar laser ranging and low redshift supernovae and then $H_1$, $U(1)$ and $\phi_2$.
%------------------------------------------------------------------------------------------------------------------------------------------------------------------------------------------%
\section{Constraints on scalar-tensor theories with lunar laser redshift and low redshift supernovae} \label{s2}
In the rest of the paper we adopt the value $\Omega_{m0}=0.27$. The results of this section weakly depend on the value of $\Omega_0$ but on the value of $H_0$. For this reason, we will choose for $H_0$ the WMAP recommended value $H_0=71\pm2.5$. The value of $\phi_1$ can be deduced from lunar laser ranging. Indeed, from lunar laser ranging\citep{Chi11}, we know that today $(\dot G/G)_{t_0}=(4\pm 9)\times 10^{-13}yr^{-1}$. But 
$$
\dot G/G=-(G^{-1})^./G^{-1}=(G^{-1})'G^{-1}H(1+z)
$$ 
with a prime meaning a derivative with respect to $z$. Developing $G^{-1}=\phi$ in power series, we thus find $(\dot G/G)_{t_0}=\phi_1 H_0$ in $z=0$. It follows that $\phi_1=0.0055\pm 0.012$. We will consider the plus sign for $\phi_1$ in (\ref{phi1}), and consequently the minus sign for $\phi_2$ in (\ref{phi2}), corresponding to the positive best fit value of $\phi_1$ got with lunar laser ranging.\\\\
In what follows, we are going to use these values of $H_0$ and $\phi_1$ as priors to determine $H_0$, $\phi_1$ and $w_0$ with supernovae data. For that, since $G$ evolves with time, we have to modify the magnitude-redshift relation usually used in General Relativity. Following \citep{RiaUza02}, we write
$$
m^{th}=5 \ln d_l-\frac{15}{4}\ln G^{-1}
$$
with $m^{th}$ the (theoretical) modulus distance and $d_l=c(1+z)\int_0^z H(z)^{-1}dz$ the usual luminosity-distance in a flat Universe. At small redshift, we get
$$
m^{th}=(5\ln z-\frac{15}{4}\ln \phi_0+5\ln \frac{c}{H_0})+(5-\frac{15}{4}\frac{\phi_1}{\phi_0}-\frac{5}{2}\frac{H_1}{H_0})z+O(2)
$$
Since $\phi_0=1$ and considering the form (\ref{H1}) for $H_1$, this rewrites as
$$
m^{th}=(5\ln z+5\ln \frac{c}{H_0})+(5-\frac{15}{4}\phi_1+\frac{5}{4}\left[-3+3w_0(\Omega_{m0}-1)\right])z+O(2)
$$
Let us remark that $\phi_1$ and $w_0$ both appear in the linear term. It means that supernovae data can only constrain a combination of these parameters and not each of them separately. The bound on $\phi_1$ got with lunar laser ranging is thus necessary to constrain efficiently $w_0$ with supernovae data. To proceed, we use the standard $\chi^2$ minimisation with some priors on $\phi_1$ and $H_0$, i.e.
$$
\chi^2=\sum_{p=1}^{n}\frac{(m^{obs}_i-m^{th})^2}{\sigma^2_i}+\frac{(\phi_1-0.0055)^2}{0.012^2}+\frac{(H_0-71)^2}{2.5^2}
$$
$n$ is the number of supernovae and $m^{obs}$ their observed distance modulus. We consider the $166$ supernovae with $z\leq 0.1$ from the last Union data\citep{Ama10}.\\
At $1\sigma$, one finds that $H_0=69.4\pm 1.2$, $\phi_1=0.0055\pm 0.0187$ and $w_0=-0.53\pm 0.68$. We got similar results for $\phi_1$ and $w_0$ when marginalising $H_0$ instead of considering it as a prior. Let us also remark that the $\Lambda CDM$ model defined by $\phi_1=0$ and $w_0=-1$ is in agreement with the data. From these values, we also derive from (\ref{phi1}) that 
$$
U(1)=(2.20\pm 0.06) 10^{-35}s^{-2}
$$
This value of the potential today ($U(1)=U(\phi_0)$) is of the same order as the cosmological constant. From (\ref{H1}) we then deduce that
$$
H_1=(2.06\pm 1.67)\times 10^{-18}s^{-1}
$$
and finally from (\ref{phi2}) that
\begin{eqnarray*}
\phi_2&=&0.50-9.04\times 10^{31}U_\phi(1)\pm\\
&&(0.54+1.15\times 10^{31}U_\phi(1)+9.43\times 10^{64}U_\phi(1)^2+\\
&&8.17\times 10^{63}\sigma_{U_\phi(1)}^2)^{1/2}s^{-2}\\
\end{eqnarray*}
with $\sigma_{U_\phi(1)}$ the error on $U_\phi(1)$. Note that all the values derived in this section are independent on the form of $U(\phi)$ but $\phi_2$. In the next section, we discuss about these results.
%------------------------------------------------------------------------------------------------------------------------------------------------------------------------------------------%
\section{Discussion} \label{s3}
In this paper we study a class of scalar-tensor theories defined by $G^{-1}=\phi$ and $U=U(\phi)$. Considering low redshift data allows then to get some constraints on the scalar field (i.e. $w_0$, $\phi_1$, $U(1)$ and a constraint on the variation of the Hubble function with $H_1$) independently on the form of its potential $U(\phi)$ by using lunar laser ranging and supernovae data. These constraints for this Brans-Dicke-like scalar field considered as a dark energy thus come from cosmology, at scales much larger than solar system scales.\\
We note that low redshift supernovae tend to lower the value of $H_0$ recommended by WMAP although it stays in the $1\sigma$ confidence contour of this last experiment. The positive value of $H_1$ shows that the Hubble function is decreasing today. Lunar laser ranging determine the value of $\phi_1$ and this allows to constrain $w_0$ with the supernovae data. Without lunar laser ranging, supernovae data only constrain a combination of $\phi_1$ and $w_0$. Constraints on $w_0$ are rather weak with respect to what we usually get when one considers high redshift supernovae but future project like JDEM should improve this situation. The constraints we got on $w_0$ and $\phi_1$ are in agreement with the $\Lambda$CDM model (for which $w_0=-1$ and $\phi_1=0$). $U(1)$ is thus accordingly close to the cosmological constant value of the standard $\Lambda$CDM model. Constraints on $\phi_2$ depend on the form of the scalar field potential.\\
Let us apply the above results to some specific scalar-tensor theories. The direct generalisation of the $\Lambda$CDM model for a scalar-tensor theory corresponds to $U=\Lambda$\citep{Mas83}. For this model, we get $\Lambda=U(1)=(2.20\pm 0.06) 10^{-35}s^{-2}$ and $\phi_2=0.50\pm 0.74$. This last range of values of $\phi_2$ is in agreement with the standard $\Lambda$CDM model since it contains the value $\phi_2=0$. Another interesting model is based on a varying potential $U=1/2m^2\phi^2$ where $m$ can be interpreted as a mass term\citep{Lin05}. The constraint on $U(1)$ allows to deduce that $m^2=2U(1)=(4.4\pm 0.12)10^{-35}s^{-2}$ and then $\phi_2=0.50\pm 0.74$. This value of $\phi_2$ is similar to the previous model, showing its robustness despite the possible variation of $U$. It could change if we consider a model with two free parameters like $U=m e^{\Lambda\phi}$\citep{Hal86}. Then, we get $m e^{\Lambda}=U(1)=(2.20\pm 0.06) 10^{-35}s^{-2}$ and $\phi_2=0.50-0.0019\Lambda\pm\sqrt{0.54+0.25\times 10^{-3} \Lambda+0.45\times 10^{-4}\Lambda^2}$. Here we can only constrain a combination of parameters i.e. $\Lambda$ and $m$. But if $\Lambda>>10^2$, $\phi_2$ has still the same value as with the two previous models and is still in agreement with a $\Lambda$CDM model.\\
The method of this paper can be applied to other scalar-tensor theories with various forms of $G$. A larger number of low redshift supernovae and a better determination of $H_0$ should allow to improve the constraints we got and to exclude or not a variation of the scalar field via the determination of $w_0$, $\phi_1$ and $\phi_2$.

\end{document}